\documentstyle[12pt,A4wide]{article}
\begin{document}
\rightline{May the 11th, 1992}
\vspace{1cm}
{\begin{center} {\Large
Thermal Reservoir coupled to External \\
\vspace{0.3cm}
Field and Quantum Dissipation} \\
\vspace{1.2cm}
{\large Fabrizio Illuminati}\footnote{Bitnet address:
illuminati@padova.infn.it} \\
\vspace{0.1cm}
{\em Dipartimento di Fisica ``Galileo Galilei",
Universit\`{a} \\
di Padova, Via F.Marzolo 8, 35131 Padova, Italia } \\
\vspace{0.3cm}
{\large Marco Patriarca}\footnote{Bitnet address:
patriarca@vaxpg.infn.it}\\
\vspace{0.1cm}
{\em Dipartimento di Fisica, Universit\`{a} di Perugia, \\
Via A. Pascoli, 06100 Perugia, Italia}
\end{center}
\vspace{1cm}
\begin{abstract}
In the framework of the Caldeira~-Leggett model of
dissipative quantum
mechanics, we investigate the effects of the interaction
of the thermal
reservoir with an external field. In particular, we
discuss how the
interaction modifies the conservative dynamics of
the central particle, and
the mechanism of dissipation. We briefly comment
on possible observable consequencies.
\end{abstract}

\vspace{0.8cm}

\vfill
DFPD 92/TH/27 \hfill May 1992
\newpage
{\large \bf 1.Introduction.}
\vspace{0.5cm}

By quantum dissipation we conventionally denote the problem of
the quantum mechanics of a particle in contact with a dissipative
environment. After pioneering work by Feynman and Vernon [1], an
exactly solvable lagrangian was put forward by Ullersma [2] in the
slightly different context of quantum brownian motion. Ullersma
introduced a mechanical model for the thermal reservoir as a set of
independent harmonic oscillators; he then showed that in the limit
of an infinite number of oscillators and of continously distributed
frequencies the dynamics of the particle is governed by the damped
Langevin equation.  Caldeira and Leggett [3] addressed the problem
of the effect of dissipation on quantum tunnelling by applying the
formalism of the effective action in the
framework of the Feynman path integral quantization.
In their analysis they
considered systems subject to both dissipative
and ``true" external forces,
thereby ruled in the quasiclassical region
by the damped equation of motion

\begin{equation}
M\ddot{q} + \eta\dot{q} + \frac{dV}{dq} = F_{ext}(t).
\end{equation}

\noindent They studied in detail the problem of the
modification of tunnelling rates in the presence of
dissipation and pioneered a new broad
field of research, mesoscopic physics,
where one is considering situations when
both thermal and quantum fluctuations
are important. For recent comprehensive
reviews on quantum brownian motion,
dissipative quantum tunnelling, and mesoscopic
physics see references [4]-[7].
The problem of quantum brownian motion
and the Ullersma~-Caldeira~-Leggett
model of quantum dissipation have also
been recently reformulated [8] in
the framework of stochastic mechanics
at finite temperature [9].
In this letter we address the question
of how the coupling of an external
field with the thermal reservoir affects
the dynamics of a dissipative quantum system.
\vspace{0.5cm}

{\large \bf 2.The model without field-reservoir coupling.}
\vspace{0.5cm}

We will handle the problem following the effective
action approach
of Caldeira and Leggett. We start by briefly
reviewing the main features of
the model without an external field interacting with
the thermal reservoir:
for a system of $N+1$ particles consider the lagrangian

\begin{equation}
L_{0} = \frac{M}{2}{\dot{q}}^{2} - V(q) + \sum_{\alpha=1}^{N}\{
\frac{m_{\alpha}}{2}{\dot{x}_{\alpha}}^{2} -
\frac{m_{\alpha}}{2}\omega_{\alpha}^{2}x_{\alpha}^{2} +
{\epsilon}_{\alpha}x_{\alpha}q \} ,
\end{equation}

\noindent which describes a particle (the so-called central
particle) of mass $M$ and potential
energy $V(q)$ interacting linearly with an
ensemble (the thermal reservoir) of
$N$ independent harmonic oscillators of
masses $m_{\alpha}$ and frequencies
${\omega}_{\alpha}$. Irreversibility and dissipation
are introduced by taking
the limit of an infinite number of thermal
oscillators with continously
distributed frequencies. The form of
the potential $V(q)$ needs not to be specified
for our purposes. By letting
$V(q) = \frac{1}{2}{{\omega}_{0}}^{2}{q}^{2}$ we
recover the Ullersma lagrangian.
Choosing a smooth function with a single metastable
minimum we recover
the Caldeira-Leggett model for dissipative quantum
tunnelling. The reduction
of the degrees of freedom of the thermal reservoir leads
to the effective
dissipative dynamics for the central particle; in
the framework of the
Feynman path integral quantization the effective
action $S_{0 \it eff}$
obtained from the euclidean version (with $q(t) = q(t + T)$ )
of the lagrangian $L_{0}$ reads

\begin{eqnarray}
S_{0 \it eff}[q(t)] & = & \int_{0}^{T}
dt(\frac{M}{2}{\dot{q}}^{2}
+ V(q)) + \nonumber \\
&   & \sum_{\alpha}
\frac{\epsilon_{\alpha}^{2}}{4m_{\alpha}
\omega_{\alpha}^{2}}\int_{0}^{T}
dt\int_{-\infty}^{\infty} dt'e^{-
\omega_{\alpha}|t-t'|}q(t)q(t') +
\nonumber \\
&   & \sum_{\alpha}\int_{0}^{T}
dt\frac{\epsilon_{\alpha}^{2}}{2m_{\alpha}
\omega_{\alpha}^{2}}q^{2}(t) ,
\end{eqnarray}

\noindent where the last term is introduced to
ensure that the coupling with the
thermal oscillators does not lower $V(q)$ below the
original uncoupled value.
Recognizing the non local time-dependent term in the action as the
dissipative force leads to the phenomenological equation (1) with
the friction coefficient $\eta$ identified as

\begin{equation}
{\eta}(\omega) = \frac{\pi}{2}\sum_{\alpha}
\frac{\epsilon_{\alpha}^{2}}{m_{\alpha}
\omega_{\alpha}^{2}}\delta(\omega - \omega_{\alpha}) ,
\end{equation}

\noindent where $\omega$ is the Fourier conjugated
of $t$ for the dynamic evolution of
the dissipative system. The above identification is
valid as long as
$\omega~\ll~\omega_{c},$ where $\omega_{c}$ is the frequency at
which the validity of
eqn.(1) begins to break down. As the number of the
thermal oscillators grows
indefinetely, the frequencies become continously
distributed according
to a spectral density $J(\omega)$ given by

\begin{equation}
J(\omega) = \frac{\pi}{2}\sum_{\alpha}
\frac{\epsilon_{\alpha}^{2}}{m_{\alpha}
\omega_{\alpha}}{\delta}(\omega - \omega_{\alpha}) .
\end{equation}

\noindent Comparing eqn.(4) and eqn.(5) one has

\begin{equation}
J(\omega) = {\eta}{\omega} .
\end{equation}

Relation (6) is a general consequence of the
requirement that the
particle-reservoir coupling, i.e. the
dissipation, be strictly linear.

\vspace{0.5cm}

{\large \bf 3.Field-reservoir interaction.}
\vspace{0.5cm}

We now want to investigate how the picture sketched
above is modified by
letting an external force act on the central
particle {\em and} on the
thermal reservoir. We are motivated by the
consideration that in many physically
interesting situations the approximation of
considering only the
central particle affected by a deterministic external
force cannot be soundly
justified. Also, from a conceptual point of view it seems
more satisfactory
to treat the central particle and the oscillators on
equal dynamical footing.
It is clear that for a generic type of coupling this
might become a formidable
problem. As a first step in our investigation we introduce a
time- and velocity-independent coupling of the
system (central particle +
oscillators) with an external field of
strength $F{_0}$, and consider a dependence at
most quadratic on the
coordinates of the system. The new model lagrangian reads

\begin{eqnarray}
L & = & \frac{M}{2}{\dot{q}}^{2} - V(q) +
F_{0}({\delta}_{0}q -
\frac{\delta'_{0}}{2}q^{2}) + \nonumber \\
  &   & \sum_{\alpha=1}^{N}\{
\frac{m_{\alpha}}{2}\dot{x}_{\alpha}^{2} -
\frac{m_{\alpha}}{2}\omega_{\alpha}^{2}x_{\alpha}^{2} +
{\epsilon}_{\alpha}x_{\alpha}q +
F_{0}({\delta}_{\alpha}x_{\alpha} -
\frac{\delta'_{\alpha}}{2}x_{\alpha}^{2}) \} .
\end{eqnarray}

We now introduce the new variables $y_{\alpha}$ defined by

\begin{equation}
y_{\alpha} = x_{\alpha} -
\frac{\delta_{\alpha}F_{0}}{m_{\alpha}\Omega_{\alpha}^{2}}
= x_{\alpha} - {\bar{x}}_{\alpha} ,
\end{equation}

\noindent with the new frequencies $\Omega_{\alpha}$
defined by

\begin{equation}
\Omega_{\alpha} = \sqrt{\omega_{\alpha}^{2} +
\frac{F_{0}\delta'_{\alpha}}{m_{\alpha}}} ,
\end{equation}

\noindent and finally define the potential $V_{F}(q)$ as

\begin{equation}
V_{F}(q) =  V(q) - F_{0}(\delta_{0}q -
\frac{\delta'_{0}}{2}q^{2}) -
\sum_{\alpha}\epsilon_{\alpha}{\bar{x}}_{\alpha}q.
\end{equation}

\noindent It is then easy to show, after some straightforward
transformations, that the lagrangian $L$ takes the form

\begin{equation}
L = \frac{M}{2}{\dot{q}}^{2} - V_{F}(q) + \sum_{\alpha}
\{\frac{m_{\alpha}}{2}\dot{y}_{\alpha}^{2} -
\frac{m_{\alpha}}{2}\Omega_{\alpha}^{2}y_{\alpha}^{2} +
\epsilon_{\alpha}y_{\alpha}q \} + E_{0}.
\end{equation}

We have thus recasted the lagrangian for the central
particle and the
oscillators interacting with the external field
in the form of the Ullersma
lagrangian for the central particle and
the oscillators alone.
The effect of the interaction
amounts now to a modification of the dynamical
parameters in the original
non-interacting lagrangian. An obvious
consequence, eqn.(9), is that
the oscillators' frequencies acquire a
field-dependent renormalization.
The interesting new feature is expressed
by eqn.(10): in the presence of an external field
interacting with the reservoir the conservative
force acting on the central
particle acquires not only a contribution from
the field-particle coupling,
but also an extra term depending on the details
of the field-reservoir
coupling. The last term in eqn. (11) defined by

\begin{equation}
E_{0} = \sum_{\alpha}
\frac{\delta_{\alpha}^{2}F_{0}^{2}}{
2m_{\alpha}\Omega_{\alpha}^{2}} =
\frac{1}{2}
\sum_{\alpha}m_{\alpha}\Omega_{\alpha}^{2}{\bar{x}}_{\alpha}^{2},
\end{equation}

\noindent amounts to a constant shift of the zero-point
energy of the
unperturbed system of thermal oscillators, and
is irrelevant to our discussion.
How the interaction of the
thermal reservoir with the external field
affects the mechanism of dissipation
is the question we want to address next. We then
consider the effective action $S_{\it eff}$
for the lagrangian $L$ defined by eqn. (11).
Since $L$ is still of the form
of an Ullersma lagrangian, we can carry over
the reduction procedure of
Caldeira and Leggett exactly as in the unperturbed case
to obtain the following expression for $S_{\it eff}$

\begin{eqnarray}
S_{\it eff}[q(t)] & = & \int_{0}^{T}
dt(\frac{M}{2}{\dot q}^{2} + V(q) -
F_{0}(\delta_{0}q - \frac{\delta'_{0}}{2}q^{2}) -
\sum_{\alpha}\epsilon_{\alpha}{\bar{x}}_{\alpha}q) + \nonumber \\
&   & \sum_{\alpha}\frac{
\epsilon_{\alpha}^{2}}{4m_{\alpha}\Omega_{\alpha}^{2}}\int_{0}^{T}
dt\int_{-\infty}^{\infty}
dt'e^{-\Omega_{\alpha}|t-t'|}q(t)q(t') + \nonumber \\
&   & \sum_{\alpha}\int_{0}^{T}dt
\frac{\epsilon_{\alpha}^{2}}{2m_{\alpha}\Omega_{\alpha}^{2}}
q^{2}(t).
\end{eqnarray}

By comparing eqn. (13) with eqns. (3), (4),
and (6) we obtain the following relations
\begin{equation}
\eta(\omega,F_{0}) = \frac{\pi}{2}\sum_{\alpha}
\frac{\epsilon_{\alpha}^{2}}{m_{\alpha}
\Omega_{\alpha}^{2}}\delta(\omega - \Omega_{\alpha}),
\end{equation}

\noindent and
\begin{equation}
J(\omega,F_{0}) = \eta(\omega,F_{0})\omega.
\end{equation}

We then recognize that the dissipation
coefficient $\eta$ and the spectral
density $J$ acquire a dependence on the
external field's strength $F_{0}$
through the renormalized frequencies $\Omega_{\alpha}$
of the thermal oscillators.
{}From the second term of eqn. (13) we see that
if $\Omega_{\alpha}$ increases with
$F_{0}$ the correlation $\langle q(t)q(t') \rangle$
is reduced; for large
times and values of $F_{0}$ one gets the locally
overdamped behaviour of the
quasiclassical equation of motion, i.e.

\begin{equation}
\dot{q} = \frac{F_{0}}{\eta} \equiv \mu{F_{0}},
\end{equation}

\noindent where $\mu = 1/{\eta}$ can be thought as the
mobility of the central particle.
The above prediction could then be tested on
microscopic models of concrete
physical systems, for instance by studying how
an external electric field
applied to an electrolitic solution (where the
overdamped central particles
are the conducting ions) affects the mobility (the friction).
\newpage
{\large \bf 4.Discussion and conclusions.}
\vspace{0.5cm}

To summarize, we have shown that the modification of the
Ullersma-Caldeira-Leggett model of quantum
dissipation to allow interaction
of the thermal reservoir with external forces
leads to some non trivial effects.
In fact, besides affecting the mechanism of
dissipation, the field-reservoir
coupling induces a new potential in the
conservative dynamics of the
central particle, depending on the field
strength {\em and} on the
dynamical parameters of the reservoir. The question
immediately arises of
where to look for such effects. A possibility is that
suggested in discussing
the consequencies of eqn. (16). However, the general
strategy should be to
apply the above scheme to explicit microscopic
models (including models of fermionic reservoirs).
In this way one could also analyse the explicit
functional
dependence of $\eta$ on the external field, and
investigate the existence of
discontinuities and phase transitions in its
behaviour for some critical
value of the field. Work is also in progress
to extend the analysis
developed in the present paper to the case
of fields with more general
dependence on the system's degrees of
freedom, and to the case of
time-dependent fields, i.e. to the problem
of the non equilibrium.

The present work is a by-product of a common
collaboration started by Pasquale
Sodano and the authors on memory effects in
quantum field theory in the
presence of external magnetic fields. We wish
to thank Prof. Sodano for
inspiring and constantly stimulating our
efforts. We also gratefully
aknowledge useful discussions with Amir Caldeira
and Silvio De Siena.
\newpage
{\begin{verse} {\Large {\bf References}} \\
\vspace{1.4cm}
[1] R.P. Feynman, and F.L. Vernon, Ann. of Phys.,
{\bf 24}, 118 (1963). \\
\vspace{0.8cm}
[2] P.Ullersma, Physica, {\bf 32}, 27 (1966). \\
\vspace{0.8cm}
[3]A.O. Caldeira, and A.J. Leggett, Ann. of Phys.,
{\bf 149}, 374 (1983). \\
\vspace{0.8cm}
[4] H. Dekker, Phys. Rep., {\bf 80}, 1 (1981). \\
\vspace{0.8cm}
[5] H. Grabert, P. Schramm, and G.L. Ingold, Phys.
Rep., {\bf 168}, 115 (1988). \\
\vspace{0.8cm}
[6] M. Razavy, and A. Pimpale, Phys. Rep.,
{\bf 168}, 305 (1988). \\
\vspace{0.8cm}
[7] G. Schoen, and A.D. Zaikin, Phys. Rep.,
{\bf 198}, 237 (1990). \\
\vspace{0.8cm}
[8] S. De Martino, S. De Siena, P. Ruggiero, and
M. Zannetti, J. Phys. A:
Math. Gen., {\bf 22}, 2521 (1989). \\
\vspace{0.8cm}
[9] P. Ruggiero, and M.Zannetti, Riv. Nuovo
Cimento, {\bf 8}, 1 (1985).
\end{verse}}
\end{document}